\documentclass[sigconf]{acmart}

\usepackage[T1]{fontenc}
\usepackage[table]{xcolor}
\usepackage{graphicx}
\usepackage{makecell} 
\usepackage{rotating} 
\usepackage{multirow} 
\usepackage{siunitx}
\usepackage{transparent}
\usepackage{url}
\usepackage[hyphenbreaks]{breakurl}
\usepackage{hyperref}

\usepackage{graphicx}
\usepackage{booktabs}
\usepackage{flushend}
\usepackage{enumitem}
\usepackage{tcolorbox}

\newcommand{\tightcolorbox}[2]{%
  {\setlength{\fboxsep}{0.5pt}\colorbox{#1}{#2}}%
}

\definecolor{DOC}{HTML}{d17d00}
\definecolor{AitM}{HTML}{ffd700}
\definecolor{LOV}{HTML}{220a7e}
\definecolor{DOS}{HTML}{ff8670}
\definecolor{UM-RM}{HTML}{e5a0be}
\definecolor{MOV}{HTML}{b832e5}
\definecolor{FP}{HTML}{ED6C52}

\begin{document}

\title{
    Poster: Towards Selecting Threat Appropriate \\ Industrial Intrusion Detection Systems
}

\author{Stefan Lenz}
\orcid{0000-0002-6486-6781}
\email{lenz@spice.rwth-aachen.de}
\affiliation{
    \institution{RWTH Aachen University}
    \city{Aachen}
    \country{Germany}
}

\author{Johannes Weidmann}
\orcid{0009-0008-7095-4332}
\email{weidmann@spice.rwth-aachen.de}
\affiliation{
    \institution{RWTH Aachen University}
    \city{Aachen}
    \country{Germany}
}

\author{Martin Henze}
\email{henze@spice.rwth-aachen.de}
\orcid{0000-0001-8717-2523}
\affiliation{
        \institution{RWTH Aachen University} 
        \city{Aachen}
        \country{Germany}
}
\additionalaffiliation{
    \institution{Fraunhofer FKIE}
    \city{Wachtberg}
    \country{Germany}
}

\setcopyright{rightsretained}
\copyrightyear{2026}
\acmYear{2026}
\setcopyright{cc}
\setcctype{by}
\acmConference[CCS '26]{Proceedings of the 2026 ACM SIGSAC Conference on Computer and Communications Security}{November 15--19, 2026}{The Hague, Netherlands}
\acmBooktitle{Proceedings of the 2026 ACM SIGSAC Conference on Computer and Communications Security (CCS '26), November 15--19, 2026, The Hague, Netherlands}
\acmDOI{10.1145/3830454.3846437}
\acmISBN{979-8-4007-2871-6/2026/11}
\begin{abstract}
    As the threat landscape against industrial control systems is dynamic, effective security requires detection strategies capable of timely reactions to these evolving threat situations.
    To address this problem, we propose the idea of a counter-threat intelligence sharing based mechanism to select appropriate detectors for current circumstances.
    To highlight the potential of this mechanism, we conduct attack-level performance evaluations of various intrusion detection systems.
    Results show the variance of intrusion detection performance depending on the attack scenario, emphasizing the benefits of such a mechanism for industrial control system security.  
\end{abstract}



\maketitle

\section{Motivation}

Industrial control systems (ICSs) become evermore interconnected and present high-value targets for adversaries facing increasingly sophisticated attacks (e.g., Stuxnet).
Since ICS are famously hard to secure, e.g., due to decade-long device-lifetimes~\cite{knapp2024_CH3}, retro-fittable security measures such as intrusion detection became popular for ICS~\cite{wolsing2022ipal}.
Consequently, a large body of research on novel, and well-performing detection approaches has been published.

As ICS exhibit predictable and deterministic behavior, attack detection for such systems is typically anomaly-based.
Thus, detectors leverage features such as network timings (e.g.,~\cite{lin2018IaT}) or the state of the physical process (e.g,~\cite{wolsing2025geco}) to detect deviations from known, benign behavior.
To test and develop novel detection mechanisms researchers typically rely on benchmarking datasets~\cite{conti2021survey} and use metrics such as recall or F1-Score~\cite{lamberts2023evaluations}.
However, as these metrics present average performance across the complete dataset and modern datasets typically contain diverse behavior and attack scenarios~\cite{conti2021survey}, such ``dataset-level'' metrics average-out performance in specific threat situations.
Furthermore, as different detection approaches leverage different features, individual detectors might be suited to detect certain attacks better.
For example, a timing-based detector could detect denial of service (DoS) attacks (e.g., flooding) quite well, while a detector using process states may not.

As effective defense strategies benefit from a holistic approach towards intrusion detection~\cite{lenz2026challenges}, related work attempts to solve this issue by combining multiple detectors with different strengths into so-called ensembles~\cite{wolsing2023ensemble,gao2021ensemble}. 
However, these again try to improve dataset-level detection, not optimizing according to the currently observed threat landscape.
In principle however, ICS operators could leverage cyber threat intelligence (CTI) sharing to make such threat specific adaptions.
For example, if an ICS operator learns from other operators, that DoS attacks are currently prevalent, they could select the most suitable detector for this threat and thus improve their security.

Therefore, in this paper we present our idea of such CTI-sharing based mechanism for dynamic intrusion detection system (IDS) selection and report our initial results. 
To this end, we present the idea behind our selection mechanism (§\ref{sec:mechanism}) including initial results showing the potential of the approach (§\ref{sec:eval}).
Then, we discuss potential challenges of realizing the selection mechanism (§\ref{sec:discussion})

\section{Threat Appropriate IDS Selection}
\label{sec:mechanism}

\begin{figure}
    \includegraphics{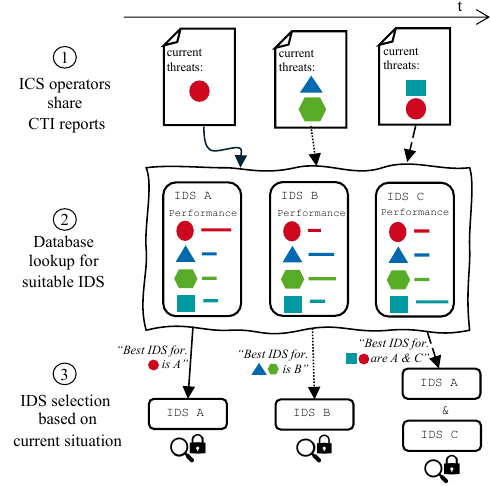}
    \caption{Sharing CTI reports among ICS operators allows selecting appropriate IDS for the current threat landscape.}
    \label{fig:design}
\end{figure}

Intrusion detection research and development currently focusses on dataset-level performance~\cite{conti2021survey,lamberts2023evaluations} obfuscating scenario-specific detection performance.
However, through sharing information about ongoing threats, organizations could optimize their detection by selecting the most suitable IDS for the current situation.
To this end, we present our idea for such a CTI-sharing based selection mechanism for (industrial) intrusion detection systems.
Broadly, this mechanism consists of three parts (shown by Fig.~\ref{fig:design}): (i) CTI analysis and sharing, (ii) an IDS performance database for individual attacker techniques, and (iii) IDS selection by cross-referencing CTI reports with the database.
In the following, we motivate the potential benefits of such a mechanism in detail.

\noindent\textbf{Step 1 - CTI Sharing:}
By increasing situational awareness, CTI sharing is widely considered to benefit (cyber-)security operations in IT systems and ICS~\cite{krief2026IcsCti}.
To this end, organizations share knowledge of adversary tactics, technique, and procedures (TTPs) in standardized formats as provided by, e.g., MITRE ATT\&CK®~\cite{mitreICSmatrix}.
This knowledge about currently prevalent TTPs can also serve as the decision basis for an informed adaptation of defense strategies.
In our case, enabling ICS operators to select the ``best'' detector for the current situation.
Therefore, the first step of our IDS selection mechanism consists of such a CTI sharing mechanism, where ICS operators share knowledge of detected TTPs with each other.
Then, assuming that similar threats might target their system as well, the administrators can select a fitting detection strategy (Fig.~\ref{fig:design}-1).

\noindent\textbf{Step 2 - IDS Performance Lookup:}
The second step of our selection mechanism comprises a ``database look-up'' telling the administrator which IDS works best for the current situation, i.e., currently observed TTPs (Fig.~\ref{fig:design}-2).
To this end, such a database must be created, by mapping attack descriptions of available benchmarking datasets to TTPs \emph{and} cross-reference performance evaluations of multiple detectors on these TTPs.
Additionally, maintaining and growing such a database allows for increasingly informed decisions and better IDS selection.

\noindent\textbf{Step 3 - IDS Selection:}
The last step of the mechanism contains the actual IDS selection (Fig.~\ref{fig:design}-3).
After cross-referencing CTI reports with the IDS database, the administrator chooses and deploys the ``most appropriate'' detection mechanism.
Moreover, this step also may include a feedback-loop, 
by confirming (or denying) an increase in detection of previously reported TTPs.

Although, such a mechanism does not exist (yet) in an end-to-end manner, research shows the benefits of CTI sharing~\cite{krief2026IcsCti}, which can also benefit intrusion detection.
To show how such an approach can be beneficial for intrusion detection, we evaluate TTP-level intrusion detection performance in the following.

\section{Proof of Concept}
\label{sec:eval}

To highlight the potential of a threat appropriate IDS selection mechanism, we show the variance in IDS detection performance across different TTPs.
To this end, we map each attack of the WDT~\cite{faramondi2021WDT} dataset to a MITRE ATT\&CK® TTP for ICS~\cite{mitreICSmatrix} and evaluate each detector on these scenarios.

\noindent
\textbf{Dataset Selection \& Mapping Procedure:}
As a basis for our experiments, we select the WDT dataset by Faramondi et al.~\cite{faramondi2021WDT}, since it contains well-labelled attacks against ICS communication (e.g., flooding attacks) and against the physical process (e.g., physical manipulation or adversary-in-the-middle) enabling the comparison of \emph{fundamentally different} detection mechanisms.
Furthermore, to discuss detector performance in a generalizable manner (see §\ref{sec:mechanism}), we map each attack to a TTP from the MITRE ATT\&CK® Matrix for ICS~\cite{mitreICSmatrix} using official CISA mapping guidelines~\cite{cisa2023mitremapping}.

\noindent
\textbf{Comparison \& IDS Selection:}
To conduct the comparison, we utilize the \texttt{IPAL} framework~\cite{wolsing2022ipal} and select four fundamentally different IDS approaches: two monitoring communication behavior and two monitoring the state of the physical process.
(i) The \texttt{InterArrivalTime} approach by Lin et al.~\cite{lin2018IaT}, leveraging consistent packet timings, (ii) \texttt{Kitsune}~\cite{mirsky2018kitsune}, monitoring various communication features, (iii) a comparatively simple detector utilizing support vector machines \texttt{OneClassSVM}~\cite{takashi2012OneClassSVM}, and (iv) the process state invariant-based detector \texttt{GeCo}~\cite{wolsing2025geco}.
We choose these particular detectors not only because they promise high detection performance, but also because they utilize different mechanisms \emph{and} features of the ICS for anomaly detection.

\begin{figure}
    \includegraphics{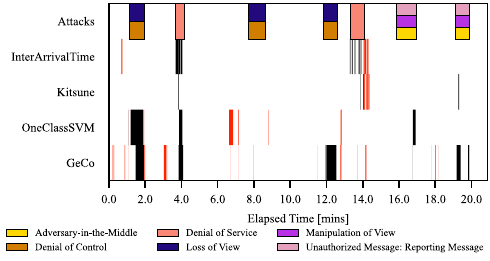}
    \caption{Labelling attacks of WDT~\cite{faramondi2021WDT} with TTPs (top row) shows how detectors alert (\tightcolorbox{black!75}{\textcolor{white}{black}} boxes = true alerts - \tightcolorbox{FP}{red} boxes = false alerts) differently based on the current TTP.
    Especially, \texttt{InterArrivalTime} performs particularly well in DoS (\tightcolorbox{DOS!50}{salmon}) scenarios, but struggles with other TTPs.
    }
    \label{fig:alerts}
\end{figure}

\noindent
\textbf{Results:}
First, mapping the attacks from WDT~\cite{faramondi2021WDT} to TTPs reveals a non-trivial challenge:
Mapping attack descriptions to TTPs does not result in clear 1-to-1 assignments.
Instead, most attack descriptions fit to multiple TTPs each describing distinct aspects of the attack.
For example, the description \emph{``Physical breakdown of pump X''} fits to the TTPs \texttt{Denial of Control} and \texttt{Loss of View} (\tightcolorbox{DOC!50}{brown} and \tightcolorbox{LOV!50}{indigo} in Fig.~\ref{fig:alerts}).
This known challenge for attack labelling~\cite{buchel2025sok}, complicates \emph{consistent} TTP assignment.
Ultimately, we assign six TTPs to three attack scenarios (Fig.~\ref{fig:alerts} - top row) covering both process-oriented TTPs (e.g., \texttt{Manipulation of View}), as well as TTPs against network components (e.g, \texttt{Denial of Service}).

To gain a high-level understanding of the behavior of each IDS across the different attack scenarios, we visualize the alerts of each detector (Fig.~\ref{fig:alerts}).
Firstly, we observe that the communication-based detectors (i.e., \texttt{InterArrivalTime}~\cite{lin2018IaT} and \texttt{Kitsune}~\cite{mirsky2018kitsune}) provide persistent alerts in the two \texttt{DoS} scenarios (\tightcolorbox{DOS!50}{salmon} in  Fig.~\ref{fig:alerts}).
However, the process-aware detectors (i.e., \texttt{OneClassSVM} and \texttt{GeCo}) do not provide reliable alerts for this TTP.
Conversely, \texttt{OneClassSVM} and \textit{GeCo} alert consistently for scenarios targeting the process state, e.g., \texttt{Adversary in the Middle} or \texttt{Manipulation of View} (\tightcolorbox{AitM!50}{gold} and \tightcolorbox{MOV!50}{magenta} in Fig.~\ref{fig:alerts}), while the communication-based IDS do not.

\begin{table}
    \begin{center}       
    \footnotesize
    \caption{The best performing detector (in \tightcolorbox{green!50}{green}) differs between attack scenarios and dataset-level evaluations do not fully capture IDS capabilities.  
    }    
    \label{tb:wdt_etar}
    \begin{tabular}{@{} l c c c c c c c @{}}
        
        \toprule
        & &  \multicolumn{6}{c}{\textbf{TTP}} \\ 
        \cmidrule(lr){3-8}
        \textbf{IDS} & \textbf{WDT}~\cite{faramondi2021WDT} & AitM & DoC & DoS & LoV & MoV & UM-RM \\
        \midrule
        InterArrivalTime~\cite{lin2018IaT} & 0.20 & 0.38 & 0.69 & \cellcolor{green!50}0.87 & 0.12 & 0.64 & \cellcolor{green!50}0.71 \\
        Kitsune~\cite{mirsky2018kitsune} & 0.07 & 0.36 & 0.36 & 0.64 & 0.09 & 0.32 & 0.25 \\
        OneClassSVM~\cite{takashi2012OneClassSVM} & 0.37 & \cellcolor{green!50}0.61 & 0.32 & 0.32 & 0.32 & 0.53 & 0.61 \\
        GeCo~\cite{wolsing2025geco} & \cellcolor{green!50}0.62 & \cellcolor{green!50}0.61 & \cellcolor{green!50}0.77 & 0.55 & \cellcolor{green!50}0.77 & \cellcolor{green!50}0.71 & 0.62 \\
        \bottomrule
    \end{tabular}
    \end{center}
    \raggedright
    \tiny\parbox{\linewidth}{\raggedright
    All scores in \texttt{eTaR}~\cite{hwang2022metrics} - higher is better. Acronyms:
    \emph{AitM}=Adversary in the Middle;
    \emph{DoC}=Denial of Control;
    \emph{DoS}=Denial of Service;
    \emph{LoV}=Loss of View;
    \emph{MoV}=Manipulation of View;
    \emph{UM-RM}=Unauthorized Message: Reporting Message.
    }
\end{table}

To solidify this understanding, we analyze the detection performance on the complete dataset and each individual TTP (Tb.~\ref{tb:wdt_etar}).
To this end, we use \texttt{eTaR}~\cite{hwang2022metrics}, a recall metric for time-series data, showing how well a detector recognizes each TTP.
On the complete dataset \texttt{GeCo} is the best performing IDS with a score of $0.62$.
However, when considering individual TTPs, e.g., \texttt{InterArrivalTime} performs substantially better in \texttt{DoS} scenarios achieving an \texttt{eTaR} of $0.87$ vs. $0.55$ of \texttt{GeCo}. 
Similarly, although \texttt{Kitsune} achieves a score of $0.09$ for \texttt{Loss of View}, its performance on \texttt{DoS} increases to $0.64$ \texttt{eTaR}, a relative difference of more than 700\%.

Overall, our results show the variance in the performance in intrusion detection based on the current attack scenario.
Furthermore, we show that utilizing TTPs for IDS evaluation enables the ``decoupling'' of performance from datasets to specific scenarios.
Ultimately, these results highlight the potential of a threat specific selection mechanism for more fine-grained intrusion detection.

\section{Future Challenges}
\label{sec:discussion}

Realizing such a selection mechanism also presents technical and non-technical challenges.
To consider these and avoid pitfalls down the road, we discuss potential problems in the following.

\noindent\textbf{CTI Sharing Incentives and Privacy Concerns:}
First, we identify the challenge of incentivizing organizations to participate in CTI sharing and not adopting the mechanism out of privacy concerns.
As any CTI-sharing approach, such problems cannot be avoided and must be addressed by providing trust of participants and guaranteeing no negative impact from sharing CTI reports~\cite{krief2026IcsCti}.
For the ICS sector, comprising a small amount of participants that are sometimes government controlled, utilizing government agencies as a trusted third party could alleviate these problems. 

\noindent\textbf{Consistent TTP Labels:}
As a technical challenge w.r.t. to creating, maintaining, and expanding the IDS database, we identify potential problems in inconsistent and unclear labelling of attacks to TTPs.
Such inconsistencies could cause a warped picture of IDS capabilities in the database leading to wrong decisions in the IDS selection phase.
However, these challenges can be minimized by utilizing automated mapping tools based on, e.g., large-language models, which show promising results for this task~\cite{buchel2025sok}.

\noindent\textbf{Timely IDS Selection and Switching:}
At last, we identify the problem of determining when to ``switch'' to another IDS for the current situation.
For example, switching as soon as new CTI reports come in might not be the best approach in all situations, since previous threats might still be relevant.
Unfortunately, simply using \emph{all} available detectors in parallel also creates uncertainty on when to report individual IDS alerts to operators~\cite{wolsing2022ipal}.
Similarly, completely switching to an IDS specialized on single TTPs might also hinder detection of other TTPs and thus potentially reducing overall detection rates.
To address such a problem, using IDS selected based on current TTPs for ensemble deployments could reduce these problems.
For example, by giving IDS for current TTPs more weight in a majority vote while still considering others to keep TTP-coverage high.

\section{Conclusion}

As threat landscapes against ICS are rapidly evolving and intrusion detectors leverage various features for anomaly detection, their performance across attack scenarios differs, too.
Leveraging this fact for better detection rates and improved ICS security, we present the idea of a CTI-sharing based IDS selection mechanism.
This mechanism cross-references the current threat landscape with an attack-scenario specific performance database and selects appropriate detectors.
Initial experiments show the variance of detection mechanisms between different attacks, highlighting the potential of  precisely selecting a well-suited detector for the current threat landscape.
Ultimately, this mechanism aims to not only improve attack detection in ICS, but also simplify and demystify detection capabilities of IDS for practical application.

%
%
%
\bibliographystyle{ACM-Reference-Format-limit}
\bibliography{references}

\end{document}